\documentclass[aps,prd,showpacs,a4paper,preprint,12pt,tightenlines]{revtex4}

\usepackage{amsmath,amsfonts,amsthm,amssymb}
\usepackage[dvips]{graphicx}
\usepackage{epsfig}
\usepackage{hyperref}
\usepackage{appendix}

\newcommand{\ud}{\mathrm{d}}

\begin{document}

\title{From model dynamics to oscillating dark energy
parameterisation}

\author{Aleksandra Kurek}
\email{kurek@oa.uj.edu.pl}
\affiliation{Astronomical Observatory, Jagiellonian University,
Orla 171, 30-244 Krak{\'o}w, Poland}
\author{Orest Hrycyna} 
\email{hrycyna@kul.lublin.pl}
\affiliation{Department of Theoretical Physics, Faculty of Philosophy, 
The John Paul II Catholic University of Lublin, Al. Rac{\l}awickie 14, 20-950
Lublin, Poland}
\author{Marek Szyd{\l}owski}
\email{uoszydlo@cyf-kr.edu.pl}
\affiliation{Astronomical Observatory, Jagiellonian University,
Orla 171, 30-244 Krak{\'o}w, Poland}
\affiliation{Mark Kac Complex Systems Research Centre, Jagiellonian University,
Reymonta 4, 30-059 Krak{\'o}w, Poland}
\date{\today}

\begin{abstract}
We develop here a relatively simple description of dark energy based on the
dynamics of non-minimally coupled to gravity phantom scalar field which, in
limit, corresponds to cosmological constant. The dark energy equation of state,
obtained directly from the dynamics of the model, turns out to be an oscillatory
function of the scale factor. This parameterisation is compared to other
possible dark energy parameterisations, among them, the most popular one, linear
in the scale factor. We use the Bayesian framework for model selection and make
a comparison in the light of SN Ia, CMB shift parameter, BAO A parameter,
observational H(z) and growth rate function data. We find that there is evidence
to favour a parameterisation with oscillations over {\it a priori} assumed
linear one.
\end{abstract}


\maketitle

\section{Introduction}

The recent discovery of the acceleration of the Universe is one of the most 
significant discoveries over last decade \cite{Riess:1998cb, Perlmutter:1998np}. 
Observations of distant supernovae type Ia \cite{Riess:1998cb, Perlmutter:1998np} 
as well as cosmic microwave background (CMB) fluctuations 
\cite{Bennett:2003bz, Spergel:2006hy} and large scale structure (LSS) 
\cite{Tegmark:2006az} indicate that the Universe is undergoing an accelerating 
phase of expansion. These observations suggest that the Universe is filled by 
dark energy of unknown form, violating the strong energy conditions
$\rho_{X} + 3p_{X} > 0$ or a dynamical equation governing gravity should be
modified. A simple cosmological constant model of dark energy can serve the
purpose of explanation of dark energy and is in good agreement with the 
astronomical data (supernovae type Ia and other measurements). Although this 
model is favoured by the Bayesian framework of model selection 
\cite{Szydlowski:2006ay,Szydlowski:2006pz, Kurek:2007tb, Kurek:2007gr}, 
it faces the serious problem of fine tuning \cite{Padmanabhan:2002ji}. 
Therefore the other alternatives \cite{Bludman:2006cg} have been proposed 
which includes an evolving scalar field. When one tries to accommodate a 
time-varying equation of state, the simplest parameterisation is the one which 
adds a linear dependence on the scale factor $a$. Other choices are motivated 
by a possibility of integration dark energy density in an exact form. In this 
context a class of simple oscillating dark energy equation of state coefficients 
appeared \cite{Linder:2005dw,Xia:2004rw, Barenboim:2004kz, Zhao:2006mn, Feng:2004ff}. 
It is interesting that these models may provide a way to unify the early 
inflation and the late time acceleration. Moreover in these scenarios we obtain 
a possible way to solve the cosmic coincidence problem
\cite{Jain:2007fa, Griest:2002cu, Dodelson:2001fq}.

If we allow that the dark energy density might vary in time (or redshift) then
there appears a problem of choosing or finding an appropriate form of parameterisation of
the equation 
of a state parameter $w_{X}(z)$. In the most popular approach
$w_{X}(z)=p_{X}/\rho_{X}$ appears to be virtual
and the dynamical dark energy
parameterisation makes the model phenomenological, containing free parameters and
functions. As a result we have a model difficult to constrain
\cite{Crittenden:2007yy}.
Another approach is to postulate a quintessence potential of the scalar field
which has motivation from fundamental physics (particle physics) and then
extracts it from the true dynamics directly \cite{Hrycyna:2007mq, Hrycyna:2007gd,
Kurek:2007bu, Faraoni:2000vg}. In this approach we can expect that the parameterisation of dark
energy equation of state reflects some realistic underlying of the
physical model. The most popular dynamical form of dark energy offers the idea of
quintessence. In this
conception dark energy is described in terms of the minimally coupled to gravity
scalar field $\phi$ with the potential $V(\phi)$. The scalar field is rolling
down its potential starts to dominate the energy density of the standard matter
\cite{Ratra:1987rm, Wetterich:1987fm}. The oscillating scalar field as a
quintessence model for dark energy has been recently proposed \cite{Dutta:2008px,
Johnson:2008se, Gu:2007be}. The case of extended quintessence introduced by
Amendola \cite{Amendola:1999dr} was also considered in our previous papers
\cite{Hrycyna:2007mq, Hrycyna:2007gd, Kurek:2007bu} where we assumed the
non-zero coupling constant. The possibility of violating of the weak energy
condition (phantom scalar field) was admitted. In this scenario, 
instead of standard minimally coupled scalar field,
the phantom scalar field,
non-minimally coupled to gravity, causes the accelerating phase of expansion of the
Universe
\cite{Hrycyna:2007mq, Hrycyna:2007gd}. We found that in
the generic case trajectories are approaching the de Sitter state after an infinite number of damping oscillations
around the mysterious $w_{X}=-1$ value. Therefore the $\Lambda$CDM model
appears as a global attractor in the phase space $(\psi,\psi')$
(where $\psi$ is a phantom scalar field and $'=\ud/\ud\ln{a}$).

In this letter, we aim at testing and selecting the viability of different
parameterisations of oscillating dark energy in the light of recent astronomical
data. We focus our attention on the equation of state for a
non-minimally coupled to gravity phantom scalar field with the potential in the simple quadratic form.

\section{Class of kinessence models}

This class of models is understood as a class of FRW models with standard dust
matter and dark
energy parameterised by redshift, i.e. $w_{X}=w_{X}(z)$ \cite{Ratra:1987rm,
Wetterich:1987fm}. For simplicity the flat FRW model ($k=0$) is assumed. Then
dynamics of the model is determined by the acceleration equation
\begin{equation}
\frac{\ddot{a}}{a}=-\frac{1}{6}\Big(\rho_{\text{m}}+(1+3w_{X})\rho_{X}\Big)=
-\frac{1}{2}H^{2}\Big(\Omega_{\text{m}}+(1+3w_{X})\Omega_{X}\Big),
\label{eq:1}
\end{equation}
where $a$ is the scale factor, dot is a differentiation with respect to the
cosmological time, $\Omega_{\text{m}}$ and $\Omega_{X}$ are density parameters for
matter and dark energy $X$, respectively, $H=(\ln{a})\dot{}$ is the Hubble
parameter.

We assume that standard matter with energy density $\rho_{\text{m}}=\rho_{\text{m},0}a^{-3}$
is a dust matter and energy density of the dark energy is given, from
conservation condition, by $\rho_{X}=\rho_{X,0}a^{-3}\exp{\Big[-3\int_{1}^{a}\frac{w_{X}(a')}{a'}\, \ud
a'\Big]}$.

The acceleration equation (\ref{eq:1}) admits the first integral (which is called
Friedman's first integral) in the form
\begin{equation}
\bigg(\frac{H}{H_{0}}\bigg)^{2} = \Omega_{\text{m},0}(1+z)^{3} + \Omega_{X,0}f(z),
\label{eq:3}
\end{equation}
where $H_{0}$ and $\Omega_{i,0}$ are parameters referring to the present epoch,
and $z$ is the redshift related to the scale factor by the relation $1+z=a^{-1}$
(the present value of the scale factor $a_{0}=1$). The phenomenological properties of dark energy are described in terms of the function $f(z)$
such that $f(z)=\exp{\Big[3\int_{0}^{z}\frac{1+w_{X}(z')}{1+z'}\, \ud z'\Big]}$.

In the context of the accelerated expansion of the Universe most
theoretical
models of dark energy are based on scalar fields. It is a consequence of exploring
an analogy to the inflationary theory of the primordial universe
\cite{Abbott:1981rg}. However, a single canonical scalar field
cannot explain the range of the coefficient of the equation of state $w<-1$ which is
preferred by the astronomical data \cite{Kurek:2007tb}. One possibility, that
has received much attention, is that we formally allow the scalar field to
have a negative kinetic energy and switch its sign in comparison with
canonical scalar field. There are some physical motivation for introducing such
a phantom scalar field arising from string/M theory and in supergravity
\cite{MersiniHoughton:2001su}. Another possibility lies in introduction of the coupling
term $\xi R\psi^{2}$ between the scalar field and the gravity (for the review
and references see \cite{Faraoni:2006ik}). Such a theory offers scalar-tensor
models of a dark energy called extended quintessence.

If we assume that a source of gravity is the phantom scalar field
$\psi$ with an arbitrary coupling constant $\xi$ then the dynamics is governed
by the action
\begin{equation}
S=\frac{1}{2}\int \ud^{4}x \sqrt{-g}\Big(m_{p}^{2}R +
(g^{\mu\nu}\psi_{\mu}\psi_{\nu} + \xi R\psi^{2} - 2 V(\psi))\Big)
\label{eq:2}
\end{equation}
where $m_{p}^{2}=(8\pi G)^{-1}$ and $V(\psi)$ is a scalar field potential. 

The phantom cosmology with general potentials was studied by Faraoni
\cite{Faraoni:2005gg} using language of qualitative analysis of differential
equation to obtain the late time attractors without the specific assumptions of
the shape of potential functions. There are many reasons why we should consider
a nonzero coupling constant $\xi$. First, a nonzero $\xi$ is generated by
quantum corrections even if it is absent in the classical action (see
\cite{Faraoni:2000gx} and references therein). Another reason is that the
non-minimal coupling is motivated by the renormalization of the
Einstein--Klein--Gordon equation. Of course the value of the coupling constant
should be fixed by the physics only, but in relativity any value of the
parameter $\xi$ different from $1/6$ (conformal coupling) gives rise to the violation
of the equivalence principle \cite{Szydlowski:2008zza}.

\begin{figure}
\includegraphics[scale=0.7]{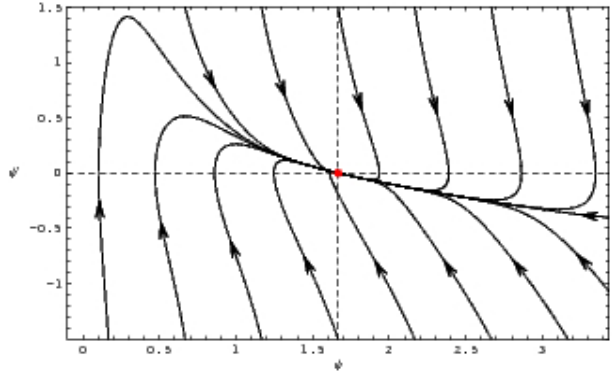}
\caption{The phase portrait represents generic behaviour of the system
(\ref{sys})
around the critical point of a stable node type.}
\label{fig:1}
\end{figure}

\begin{figure}
\includegraphics[scale=0.7]{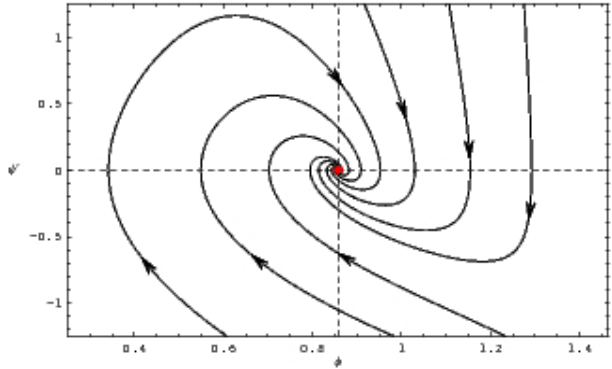}
\caption{The phase portrait represents
the generic behaviour of the system (\ref{sys}) around a focus type critical point.}
\label{fig:2}
\end{figure}

In our paper \cite{Hrycyna:2007gd} we studied generic features of the
evolutional
paths of the flat FRW model with the phantom scalar field non-minimally coupled to
gravity by using dynamical systems methods.
We reduced dynamics of the model to the simple case of autonomous
dynamical system on the invariant submanifold $(\psi,\psi')$ (a prime denotes
differentiation with respect to the natural logarithm of the scale factor) 
\begin{align}
\label{sys}
\psi' &= y \nonumber \\
y' &= -y- y^{2}(y+6\xi\psi)\frac{1-6\xi}{1+6\xi\psi^{2}(1-6\xi)} - \\
&- 
\frac{1+(1-6\xi)y^{2}+6\xi(y+\psi)^{2}}{1+6\xi\psi^{2}(1-6\xi)}
\bigg[ \frac{2\psi(y+6\xi\psi) -
\big(1+6\xi\psi(y+\psi)\big)}{\psi}\bigg]  \nonumber  
\end{align}
where $V(\psi) \propto \psi^2$ is assumed, and
we found that principally there is one asymptotic state, which corresponds to
the critical point in the phase space $\psi_{0}=\pm1/\sqrt{6\xi}$ and
$\psi_{0}'=0$. This critical point is also the de Sitter state ($w_{X}=-1$).
Note
that in this model a problem of the big rip singularity does not appear in contrast to 
the standard phantom cosmology where it is present because the late time attractors in
the phase space represent the de Sitter stage. There
are two types of evolutional scenarios leading to this Lambda state (depending
on the value of $\xi$), through
\begin{itemize}
\item[1.]{the monotonic evolution toward the critical point of a node
type for $0<\xi\le3/25$, (Fig.~\ref{fig:1}), in the special case $\xi=3/25$ we
obtain a degenerate node;}
\item[2.]{the damping oscillations around the critical point of a focus type
for $3/25<\xi<1/3$, (Fig.~\ref{fig:2}).}
\end{itemize}

The effect of a non-minimal coupling can be treated as an effect of fictitious
fluid with some effective coefficient of the equation of state given by
\begin{equation}
w_{\mathrm{eff}} = \frac{2}{1+6\xi\psi^{2}(1-6\xi)}\bigg\{\frac{1}{2}[1+2\xi\psi^{2}(1-6\xi)]
 -(1-2\xi)[1+(1-6\xi)\psi'^{2}] - 4\xi(1-3\xi)(\psi'+\psi)^{2}\bigg\}.
 \end{equation}
In both evolutional scenarios we can find linearised solutions of the dynamical
system in the vicinity of the critical point, which the following Hartman-Grobman
theorem are good approximations of the system.

Finally one can compute linearised formulas for $w(z)$ around the corresponding
critical point for both cases (see Appendix \ref{appa}).
 
In what follows we concentrate on the special case of parameterisation
(\ref{osc}) with $\xi=1/6$ which corresponds to the conformally coupled phantom
scalar field \cite{Szydlowski:2008zza}. We will confront it (using the Bayesian
model selection method) with most popular dark energy parameterisations of
$w_X(z)$, which are presented in Table~\ref{tab:1}.

\begin{table}
\caption{\label{tab:1}Different dark energy parameterisations in terms of
$w_{X}(z)\equiv p_{X}/\rho_{X}$ -- the coefficient of EoS.}
\begin{tabular}{|c|l|}
\hline
case & parameterisation  \\
\hline
(1) & Chevallier-Polarski-Linder \cite{Chevallier:2000qy,Linder:2004ng} \\
    &$w_X(z)=w_{0}+w_{1}\frac{z}{1+z}$ \\
\hline
(2) & purely oscillating dark energy  \\
    &a) $w_X(z)=w_{0}\cos{(\omega_{c}\ln{(1+z)})}$ \\
    &b) $w_X(z)=-1+w_{0}\sin{(\omega_{s}\ln{(1+z)})}$ \\
\hline
(3) & damping osc. DE \\
    &a) $w_X(z)=w_{0}(1+z)^{3}\cos{(\omega_{c}\ln(1+z))}$ \\
    &b) $w_X(z)=-1+w_{0}(1+z)^{3}\sin{(\omega_{s}\ln{(1+z)})}$ \\
\hline
(4) & damping osc. DE parameterisation determined directly from \\ 
    & the dynamics of phantom scalar field model \cite{Hrycyna:2007gd}  \\
   & a) $\xi =\frac{1}{6}$:\\   &$w_X(z)=-1-\frac{4}{3}(1+z)^{3/2}\Big((\cos{(\frac{\sqrt{7}}{2}\ln{(1+z)})} +
\frac{5\sqrt{7}}{7}\sin{(\frac{\sqrt{7}}{2}\ln{(1+z)})})x_{0}+$\\
 & $(\cos{(\frac{\sqrt{7}}{2}\ln{(1+z)})} +
\frac{\sqrt{7}}{7}\sin{(\frac{\sqrt{7}}{2}\ln{(1+z)})})y_{0}\Big)$\\
& $-\frac{2}{3}(1+z)^3\Big((\cos{(\frac{\sqrt{7}}{2}\ln{(1+z)})} +
\frac{5\sqrt{7}}{7}\sin{(\frac{\sqrt{7}}{2}\ln{(1+z)})})x_{0}+$\\
 & $(\cos{(\frac{\sqrt{7}}{2}\ln{(1+z)})} +
\frac{\sqrt{7}}{7}\sin{(\frac{\sqrt{7}}{2}\ln{(1+z)})})y_{0}\Big)^2$\\
    &b) $\xi =\frac{1}{6}$, $y_0= \alpha x_0$ : \\
&$w_X(z)=-1-\frac{4}{3}(1+z)^{3/2}\Big((1+\alpha)\cos{(\frac{\sqrt{7}}{2}\ln{(1+z)})} +
\frac{\sqrt{7}}{7}(5+\alpha)\sin{(\frac{\sqrt{7}}{2}\ln{(1+z)})}\Big)x_{0}-$\\
& $-\frac{2}{3}(1+z)^3\Big((1+\alpha)\cos{(\frac{\sqrt{7}}{2}\ln{(1+z)})} +
\frac{\sqrt{7}}{7}(5+\alpha)\sin{(\frac{\sqrt{7}}{2}\ln{(1+z)})}\Big)^{2}x_{0}^{2}$\\
\hline
\end{tabular}
\end{table}

Recently, it has been argued that models with
oscillating dark energy are favoured over a model with a linear parameterisation of
EoS \cite{Kurek:2007bu, Jain:2007fa}.  

\section{Results}
\subsection{Bayesian method of model comparison}
To find the best parameterisation of $w_X(z)$ we use the Bayesian method of model
comparison \cite{Jeffreys:1961}. Here the best model ($M$) from the set of
models under consideration is the one which has the greatest value of the
probability in the light of the data ($D$) (posterior probability)
\begin{equation}
P(M|D)=\frac{P(D|M)P(M)}{P(D)}.
\end{equation}
$P(M)$ is the prior probability for model $M$, $P(D)$ is the normalisation
constant and $P(D|M)$ is the model likelihood (also called evidence) and is
given by $P(D|M)=\int P(D|\bar{\theta},M)P(\bar{\theta}|M) d\bar{\theta}$, where $P(D|\bar{\theta},M)=\mathrm{L}(\bar{\theta})$ is the likelihood function for model
$M$ and $ P(\bar{\theta}|M)$ is the prior probability for the model parameters
$\bar{\theta}$. It is convenient to consider the ratio of the posterior probabilities for models which we want to compare $\frac{P(M_1|D)}{P(M_2|D)}=\frac{P(D|M_1)}{P(D|M_2)}\frac{P(M_1)}{P(M_2)}$. If we have no prior information to favour one model over another one
($P(M_1)=P(M_2)$), posterior ratio is reduced to the ratio of the model
likelihoods, so called Bayes factor ($B_{12}$), which values can be interpreted as the strength of evidence to favour model $M_1$ over model $M_2$ \cite{Trotta:2008qt}:
$0< \ln B_{12} <1 $ -- `inconclusive'; $1<\ln B_{12} <2.5$
-- `weak';  $2.5<\ln B_{12} <5$ -- `moderate'; $\ln B_{12} >5$ --
`strong'.

The values of Bayesian evidence for models with $w_X(z)$ defined in Table~\ref{tab:1} were obtained using a nested sampling algorithm \cite{Skilling}, which implementation to the
cosmological case is available as a part of the CosmoMC code \cite{cosmo:1,
Lewis:2002ah}, called CosmoNest \cite{cosmo:2, Mukherjee:2005wg,
Mukherjee:2005tr, Parkinson:2006ku}. It was changed for our purpose. We
assume flat prior probabilities for the model parameters in the following intervals:
$\Omega_{\text{m},0} \in [0,1]$ and $w_0 \in [-2,0]$, $w_1 \in [-3,3]$ (Model 1); $w_0 \in [-2,0]$, $\omega_c \in [0,2]$ (Model 2a and Model 3a); $w_0 \in [-2,2]$, $\omega_s \in [0,2]$ (Model 2b and Model 3b); $x_0 \in [-1,1], y_0 \in [-1, 1]$ (Model 4a); $x_0 \in [-1,1], \alpha\in[-3,0]$ (Model 4b). The values of evidence were averaged from the eight runs.

\subsection{Analysis with SNIa, CMB R and BAO A data}

To compare models gathered in Table~\ref{tab:1} we use information coming from the sample of $N_1=192$ SNIa data    
\cite{Davis:2007na}, which consists of the ESSENCE sample \cite{WoodVasey:2007jb} and a SNIa detected by HST \cite{Riess:2006fw}. After suitable calibration SNIa could be treated as standard candles and tests on the assumed cosmology could be done. In this
case the likelihood function has the following form
\begin{equation}
\mathcal{L}'_{\text{SN}}\propto \exp
\left[-\frac{1}{2}\left(\sum_{i=1}^{N_1}\frac{(\mu_{i}^{\text{theor}}-\mu_{i}^{\text{obs}})^{2}}{\sigma_{i}^{2}}\right)
 \right],
\end{equation}
where $\sigma_{i}$ is known, $\mu_{i}^{\text{obs}}=m_{i}-M$ ($m_{i}$--apparent magnitude, $M$--absolute magnitude of SNIa), $\mu_{i}^{\text{theor}}=5\log_{10}D_{Li} +
\mathcal{M}$, $\mathcal{M}=-5\log_{10}H_{0}+25$ and $D_{Li}=H_{0}d_{Li}$, where
$d_{Li}$ is the luminosity distance, which (with the assumption that the Universe is spatially flat) is given by
$d_{Li}=(1+z_{i})c\int_{0}^{z_{i}} \frac{dz'}{H(z')}$ and $H(z)$ is defined in
equation~(\ref{eq:3}). After an analytical marginalisation over the nuisance
parameter $\mathcal{M}$ one can obtain the likelihood function
$\mathcal{L}_{\text{SN}}$ which does not depend on the parameter $H_0$.

We also include information coming from the CMB data using measurement of the shift parameter ($R^{\text{obs}}=1.70 \pm 0.03$ for $z_{\text{dec}}=1089$) \cite{Spergel:2006hy,Wang:2006ts}, which is related to the first acoustic peak in the temperature power spectrum and is given by $R^{\text{theor}} =
\sqrt{\Omega_{\text{m},0}}\int_{0}^{z_{dec}}\frac{H_0}{H(z)}dz$. Here the likelihood
function has the following form 
\begin{equation}
\mathcal{L}_{R} \propto \exp \left[- \frac
 {(R^{\text{theor}}-R^{\text{obs}})^2}{2\sigma_{R}^2} \right].
\end{equation}

As the third observational data we use the SDSS luminous red galaxies measurement of $A$ parameter ($A^{\text{obs}}=0.469 \pm 0.017$ for
$z_{A}=0.35$) \cite{Eisenstein:2005su}, which is related to the baryon acoustic
oscillations (BAO) peak and defined in the following way $A^{\text{theor}}=\sqrt{\Omega_{\text{m},0}} \left (\frac{H(z_A)}{H_{0}}
\right ) ^{-\frac{1}{3}} \left [ \frac{1}{z_{A}} \int_{0}^{z_{A}}\frac
{H_0}{H(z)} dz\right]^{\frac{2}{3}}$. In this case the likelihood function has the following form 
\begin{equation}
\mathcal{L}_{A} \propto \exp \left[
 -\frac{(A^{\text{theor}}-A^{\text{obs}})^2}{2\sigma_{A}^2}\right ].
\end{equation}

The final likelihood function used in analysis is given by $\mathcal{L}=\mathcal{L}_{\text{SN}}\mathcal{L}_{R}\mathcal{L}_{A}$.
\\
The results, i.e.
values of $\ln B_{1i}$ together with their uncertainties, computed with respect to the model with linear in $a$
parameterisation of $w_X(z)$, are presented in the first column of Table~\ref{tab:2}.
 
As we can conclude there is weak evidence to favour model with purely
oscillations (2b) over the model with
linear in $a$ parameterisation. The comparison with models 2a, 3b and 4a is inconclusive, which means that the data set used in analysis is not enough powerful to distinguish those models. Additional information coming from different data set or more accurate data set is required. The value of logarithm of the Bayes factor calculated with respect to model 4b is close to $-1$, which could indicates on the weak evidence in favour of it, but more information is needed to make the conclusion more robust.  There is moderate evidence to favour model 1 over the model with damping
oscillations (3a).

We can also check if the damping term, i.e. $(1+z)^3$, is required by the data.
Comparing model 2a with 3a one can conclude that there is moderate evidence to
favour the purely oscillations ($\ln B_{\text{2a3a}}=4.32$), while comparing 2b
with 3b that there is weak evidence to favour purely oscillations ($\ln
B_{\text{2b3b}}=2.13$).

The comparison of the best
parameterisation from the set of models with purely oscillations (2b) over the
best one among the models with damping oscillations (4b) does not give conclusive answer: $\ln B_{\text{2b4b}}=1.06$.

Finally one can compare models with dynamical dark energy with parameterisations
of the equation of state gathered in Table \ref{tab:1} with the simplest
alternative, i.e. the $\Lambda$CDM model with $w=-1$. As one can conclude this
model is still the best one in the light of SNIa data, CMB $R$ shift parameter
and BAO $A$ parameter. However the conclusion from the comparison of the
$\Lambda$CDM model with the purely oscillating model (2b) is inconclusive ($\ln
B_{\Lambda\text{CDM,2b}}=1.05$). 
\\

\begin{figure}
\includegraphics[scale=0.65]{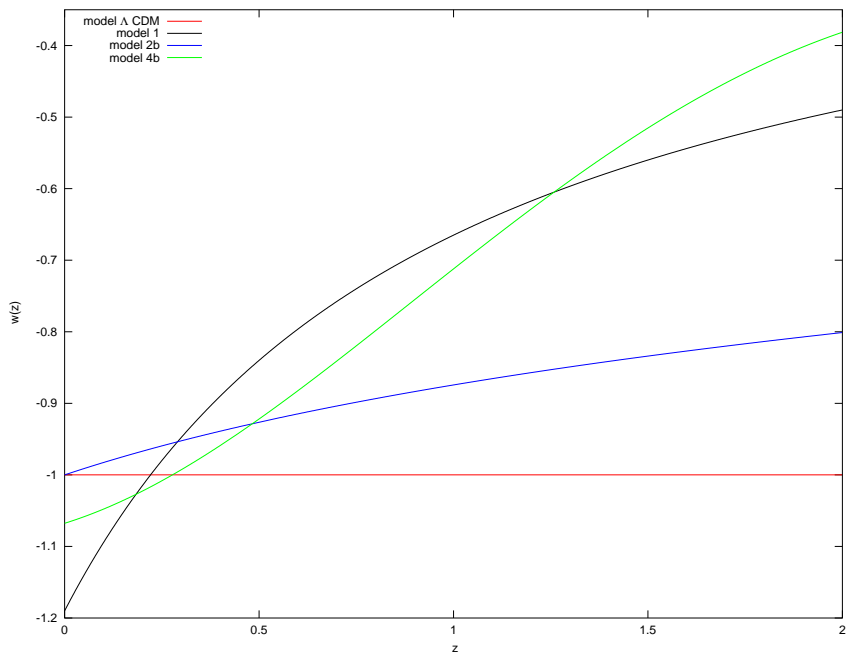}
\caption{The functions $w(z)$ for the $\Lambda$CDM model, model with linear in $a$ parameterisation of $w(z)$ (1), model with purely oscillations (2b) and model with damping oscillations (4b), calculated for the best fit values of model parameters (SNIa+CMBR+BAOA data).}
\label{fig:3}
\end{figure}

\begin{figure}
\includegraphics[scale=0.65]{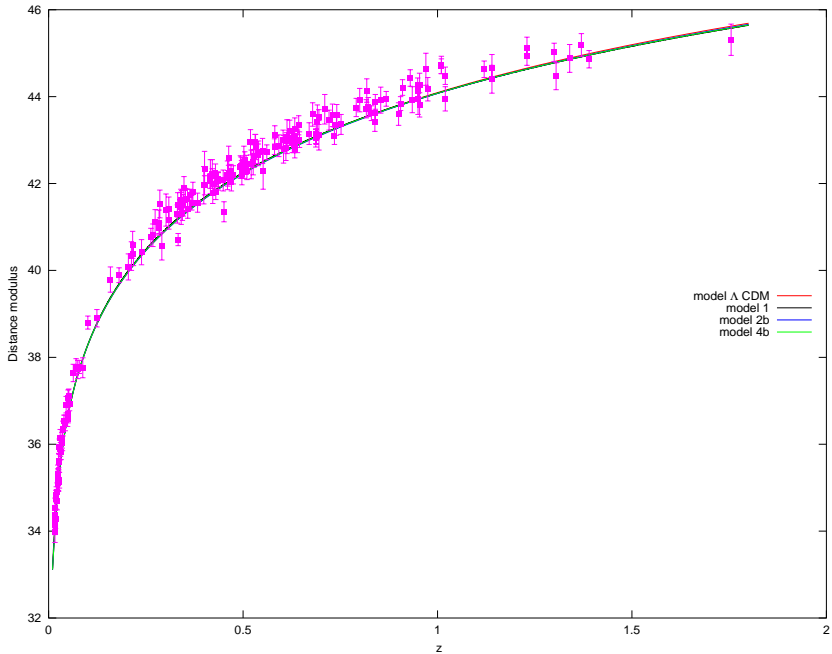}
\caption{The distance modulus vs redshift relations for the $\Lambda$CDM model, model with linear in $a$ parameterisation of $w(z)$ (1), model with purely oscillations (2b) and model with damping oscillations (4b), calculated for the best fit values of model parameters (SNIa+CMBR+BAOA data) and with the assumption that $H_0=72$ kms${}^{-1}$ Mpc${}^{-1}$. The SNIa data set is also presented.}
\label{fig:4}
\end{figure}

It is interesting that the model with purely oscillations (2b) is favoured by the data over the model with linear in $a$ parameterisation of $w(z)$. Also the model with damping oscillations (4b) fares well when compared with model 1. One can try to understand the reason of such conclusions. In Figure~\ref{fig:3} the functions $w(z)$ for the $\Lambda$CDM model, model 1,  model 2b and model 4b are presented, which were calculated for the best fit values of the model parameters (obtained in the analysis with the SNIa, CMB R and BAO A data). While in Figure~\ref{fig:4} we present the corresponding distance modulus vs redshift relations (with the additional assumption that $H_0=72$ kms${}^{-1}$ Mpc${}^{-1}$).

As we can conclude in spite of the prominent differences in the functions $w(z)$, the distance modulus relations are nearly identical for those models. One should keep in mind that parameters were fitted using the data which are based on the luminosity distance. In this case $w(z)$ is integrated twice.

\subsection{Analysis with $H(z)$ data added}
It is interesting to consider the relation $H(z)$, as it depends on the $w(z)$ through one integral. Unfortunately the present Hubble function measurements on different redshifts are small and inaccurate. However $H(z)$ data set could gives us another insight into the problem considered.

One possibility to measure the Hubble parameter as a function of redshift is based on the differential ages $dt/dz$ of passively evolving luminous red galaxies (LRG), which correspond to the Hubble function through the relation
\begin{equation}
H(z)=-\frac{1}{1+z} \frac{dz}{dt}.
\end{equation}
Using Gemini Deep Deep Survey and archival data the authors of \cite{Simon:2004tf} obtained nine values of the Hubble parameter for different redshifts in the range $0.09<z<1.75$. Although this data set is small and has large uncertainties we include it in our analysis.

Another method to determine the Hubble function values at various redshifts is based on the line of sight (LOS) baryon acoustic oscillation scale measurements. The scale of the BAO in the radial direction depends on the $H(z)$. On the other hand the precise measurement of this scale is given by the CMB observations, so the comparison gives us the value of Hubble parameter. Based on this method and using the SDSS DR6 luminous red galaxies data the authors of \cite{Gaztanaga:2008xz} obtained the values of $H$ at three different redshifts. The uncertainties are highly reduced when compared with the previous data set. We include those points in our analysis. 

To complete our $H(z)$ data set we use the HST measurement of $H_0$ \cite{Freedman:2000cf}. 

We repeat previous calculations with the additional $N_2=13$ Hubble function measurements. The corresponding likelihood function has the following form: $\mathcal{L}=\mathcal{L}_{SN}\mathcal{L}_{R}\mathcal{L}_{A}\mathcal{L}_{H}$, where
\begin{equation}
\mathcal{L}_{H} \propto \exp \left[-\frac{1}{2} \sum_{i=1}^{N_2}\left( 
 \frac{(H^{\text{theor}}(z_i)-H_i^{\text{obs}})^2}{\sigma_{H i}^2}\right ) \right ].
\end{equation}
The values of $\ln B_{1i}$ together with uncertainties are gathered in the second column of Table~\ref{tab:2}. As one can conclude the inclusion of $H(z)$ data does not change our conclusion in most cases. There is still weak evidence to favour of model with purely oscillations (2b) over the model with linear in $a$ parameterisation of $w$. Evidence in favour of model 4b becomes slightly greater (weak evidence to favour this model over the model 1). The evidence against model 3a is even greater than in previous calculations, we find strong evidence against it.
 
The evidence to favour model with purely oscillations (2a) over the model with damping term (3a) is strong ($B_{2a3a}=6.2$), while the evidence in favour of model 2b over the model 3b is moderate ($B_{2b3b}=2.57$).

The comparison of the best model among the models with purely oscillations, i.e. 2b, with the best model from the set of models with damping term, i.e. 4b, does not give the conclusive answer $B_{2b4b}=0.99$. As one can conclude the $\Lambda$CDM model is still the best one from the models considered, however the conclusion from the comparison with model 2b is inconclusive ($B_{\Lambda\text{CDM},2b}=0.97)$. 

\begin{figure}
\includegraphics[scale=0.65]{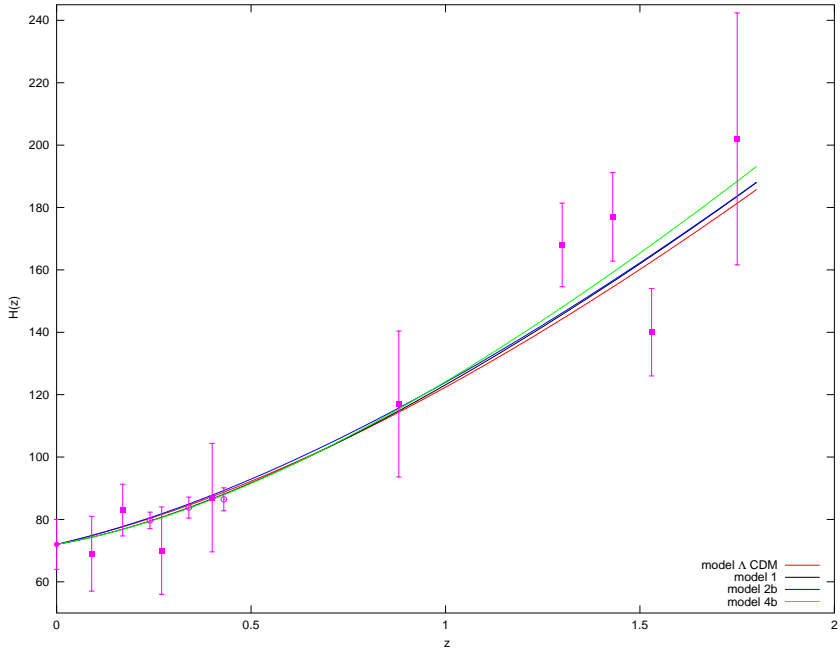}
\caption{The $H(z)$ functions for the $\Lambda$CDM model, model with linear in $a$ parameterisation of $w(z)$ (1), model with purely oscillations (2b) and model with damping oscillations (4b), calculated for the best fit values of model parameters (SNIa+CMBR+BAOA+H data). The filled square, circle and filled circle points correspond to observational $H$ data from \cite{Simon:2004tf}, \cite{Gaztanaga:2008xz} and \cite{Freedman:2000cf}, respectively.}
\label{fig:5}
\end{figure}

We present the relations $H(z)$ for model 1, model 2b, model 4b and the $\Lambda$CDM model in Figure~\ref{fig:5}. The Hubble functions were derived for the best fit model parameters in the analysis with SNIa, CMB R, BAO A and observational H(z) data.

As one conclude the relations $H(z)$ for models considered are similar in the redshift range under consideration. More data with better quality is required. The most promising future $H$ data will come from the BAO measurements. This method gives us much more precise data points when compared with the alternative method. 

\subsection{Analysis with growth rate function data added}
The conclusions stated before are based on the geometrical dark energy probes. It is interesting to check how the inclusion of the dynamical probes, related to the growth of structures, will change the results. We consider observations of the growth rate function $f$, which is related to the growth function $D$ by the following formula $f\equiv d \ln D / d \ln a $. Its evolution in the general relativity framework is described by the following equation
\begin{equation}\label{eq:f}
a\frac{df}{da}=-f^2-f\left(\frac{1}{2} -\frac{3}{2} (1-\Omega_{\text{m}}(a))w(a) \right) + \frac{3}{2}\Omega_{\text{m}}(a),
\end{equation}
where $\Omega_{\text{m}}(a)=\frac{\Omega_{\text{m},0}a^{-3}}{H^2/H_0^2}$. 

The values of growth rate ($f^{\text{theor}}$) at various scale factor ($a$) for considered models were obtained with the help of eq.~\ref{eq:f}. It was solved using numerical methods, with the assumption that $f(a \simeq 0)=1$. 

The observational growth rate data ($f^{\text{obs}}$) could be obtained through the measurements of the redshift distortion parameter $\beta$. It is observed through the anisotropic pattern of galactic redshifts on cluster scales. It is related to the growth rate function by the following formula $\beta \equiv f / b$. The so called bias parameter $b$ reflects the fact that the galaxy distribution does not perfectly trace the matter distribution in the Universe. Currently there are only few measurements of $f$ available (see Table~\ref{tab:3}). This data set is similar to the one presented in \cite{Wei:2008rv}. We do not consider the data points at $z=0.55$ and $z=1.4$, as the bias parameter was derived with the help of the value $\beta$ in those cases. The measurement at $z=3$ was obtained in different method, which does not rely on $\beta$ and $b$ parameters. The value of $f$ is finding in the analysis with Ly-$\alpha$ forest data. 

 \begin{table}
    \centering
\caption{The values of distortion parameter $\beta$, bias parameter $b$ and corresponding growth rate function $f=\beta b$ which are used in the calculations.}
    \begin{tabular}{c|c|c|c|c}
    z & $\beta$ & $b$ & $f$ & references  \\
    \hline 
     0.15 &	0.49 $\pm 0.09$ & 1.04 $\pm 0.11$& 0.51 $\pm 0.11$ & \cite{Hawkins:2002sg}, \cite{Verde:2001sf} \\
     0.35 &	0.31 $\pm 0.04$ & 2.25 $\pm 0.08$& 0.70 $\pm 0.18$ & \cite{Tegmark:2006az} \\
     0.77 &	0.70 $\pm 0.26$ & 1.30 $\pm 0.10$& 0.91 $\pm 0.36$ & \cite{Guzzo:2008ac} \\
     3.00 &	 - & - & 1.46 $\pm 0.29$ & \cite{McDonald:2004xn}\\
     \hline
\end{tabular}
\label{tab:3}
\end{table}

It should be kept in mind that this data set was obtained with the assumption of
the $\Lambda$CDM model. Its inclusion in the analysis with other models could
decrease its reliability and the results should be treated with care.

The likelihood function used in analysis is of the following form
$\mathcal{L}=\mathcal{L}_{SN}\mathcal{L}_{R}\mathcal{L}_{A}\mathcal{L}_{H}\mathcal{L}_{f}$, where
\begin{equation}
\mathcal{L}_{f} \propto \exp \left[-\frac{1}{2} \sum_{i=1}^{N_3} \left( 
 \frac{(f^{\text{theor}}(a_i)-f_i^{\text{obs}})^2}{\sigma_{fi}^2} \right) \right ],
\end{equation} 
where $N_3=4$.
The values of $\ln B_{1i}$ and its uncertainties are gathered in the third column of Table~\ref{tab:2}. 

As one can conclude the final conclusions do not change in all cases. The data set is not informative enough to change results. 

\begin{figure}
\includegraphics[scale=0.65]{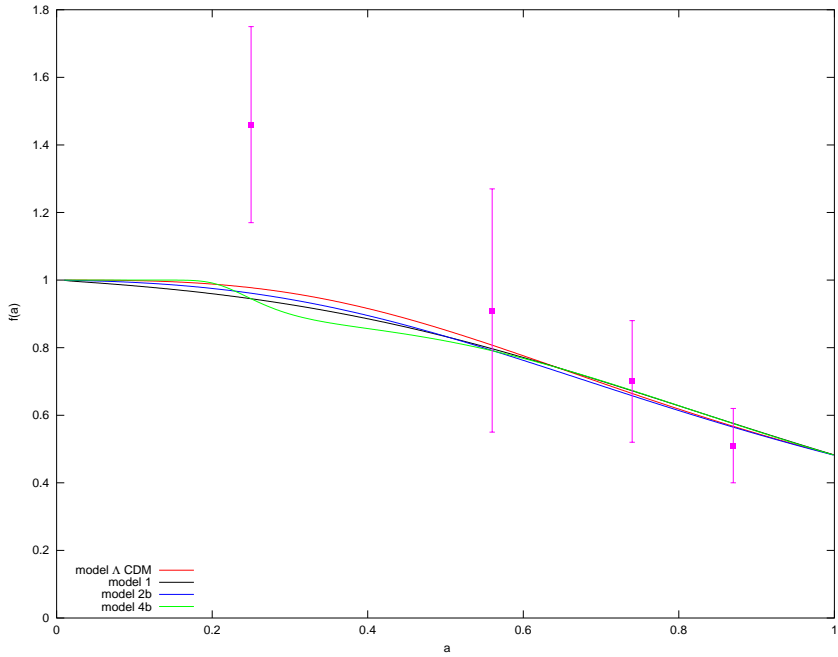}
\caption{The $f(z)$ functions for the $\Lambda$CDM model, model with linear in $a$ parameterisation of $w(z)$ (1), model with purely oscillations (2b) and model with damping oscillations (4b), calculated for the best fit values of model parameters (SNIa+CMBR+BAOA+H+f data).}
\label{fig:6}
\end{figure}

In Figure~\ref{fig:6} one can find a plot of the growth rate as a function of the scale factor for the $\Lambda$CDM model, model 1, model 2b and model 4b, calculated for the best fit values of model parameters (in the analysis with SNIa,CMB R, BAO A, H and f data).   

The relation $f(a)$ for model 4b differs from the other relations. Anyway the data points have large uncertainties, which prevent this set to distinguish models considered. This is in agreement with our previous conclusion.

 \begin{table}
    \centering
\caption{The values of $\ln(B_{1i})=\ln P(D|M_1) - \ln P(D|M_i)$ calculated with respect to the model with linear
in $a$ parameterisation of $w_X(z)$ for different data sets.}
    \begin{tabular}{c||c|c|c}
    MODEL & SNIa+CMBR+BAOA& SNIa+CMBR+BAOA+H& SNIa+CMBR+BAOA+H+f  \\
    \hline 
     1 &	0 & 0& 0\\
     2a & 0.29 $\pm 0.22$ & 0.33 $\pm 0.21$ &0.18 $\pm 0.22$ \\
     2b	& -2.08 $\pm 0.13$ & -2.14 $\pm 0.20$ & -2.24 $\pm 0.23$\\
     3a	&	4.61 $\pm 0.18$ & 6.53 $\pm 0.24$ & 6.3 $\pm 0.23$\\
     3b	&	0.05 $\pm 0.12$ &0.43 $\pm 0.23$ & 0.34 $\pm 0.25$\\
     4a  & 0.5 $\pm 0.13$ & 0.31 $\pm 0.23$ & 0.25 $\pm 0.24$\\
     4b  & -1.02 $\pm 0.18$ & -1.15 $\pm 0.23$ & -1.21 $\pm 0.23$\\
     \hline
     $\Lambda$CDM & -3.13 $\pm 0.16$ &-3.11 $\pm 0.22$ & -3.3 $\pm 0.23$\\
\end{tabular}
\label{tab:2}
\end{table}

\subsection{Discussion}

In spite of the fact that models with oscillating relation for $w(z)$ (i.e. 2b and 4b) fare well when compared with the model in which $w(a)$ is a linear function of scale factor, the oscillating behaviour is not seen in the $w(z)$ vs $z$ plots in the redshift interval considered. While the frequency parameter of the model 4b is fixed by the theory, it appears as a free parameter in model 2b. The relation for $w(z)$ in model 4b is complicated. However it can be rewritten as sum of sine and cosine components, with the amplitudes, which depend on $z$, as well as on the model parameters. On the other hand frequency parameters $w_s$ are equal to $\sqrt{7}/2$ or $\sqrt{7}$. If we consider the relation $| w_s \ln(1+z) |= 2 \pi$, we can claim that the oscillating behaviour should be observed in the redshift range of about $\Delta z \simeq 10 $. Unfortunately most of the data points used in analysis are for $z < 2 $. The oscillating behaviour is not seen. More observations at higher redshifts are needed. On the contrary, as was stated before, the frequency parameter of model 2b is a free one. The assumed prior range for this parameter (i.e. $w_s \in [0,2]$) corresponds to the period of oscillation of at least $\Delta z \simeq 22 $. It is again too big to be observed with the present data sets. It is interesting to consider the situation in which the oscillating behaviour could be detected. It can be done by assessing a different prior range for frequency parameter. We repeat calculations for model 2b, with the assumption that $w_s \in [2, 4.5]$ (model 2b1). It corresponds to period of oscillations of at least $\Delta z \simeq 3 $ (the redshift of the most distant data point, of course apart from the one at $z=1089$). The value of logarithm of the Bayes factor, calculated with respect to model 2b, is equal to $\ln B_{2b,2b1} = 4.1 \pm 0.17$. This means that the evidence against model 2b1 is moderate. We can conclude that available data sets prefer a model with period of oscillations larger than could be detected nowadays.

\section{Conclusions}
We use the Bayesian method of model selection to compare the FRW models with different
functional forms of dynamical dark energy (different parameterisations of the
EoS). We examine two categories of parameterisations: {\it a priori\/} assumed and
derived from the model dynamics. We show that two parameterisations are favoured
over most popular linear with respect to the scale factor.

In particular we obtain following results:
\begin{itemize}
    \item parameterisation with purely oscillations, i.e. 2b, is the best one among parameterisation considered in this paper;
    \item there is weak evidence to favour this parameterisation over
    the linear in $a$ parameterisation of EoS (this conclusion is based on the SNIa, CMB R and BAO A data sets and does not change after inclusion observational $H$ and $f$ data);
    \item data sets used in analysis prefer model 2b in which oscillating behaviour could not be detected nowadays 
    \item comparison of model 4b with the linear in $a$ parameterisation of EoS does not give conclusive answer when it is based on SNIa, CMB R and BAO A data, but after the inclusion of $H$ and $f$ data we find the weak evidence in favour of this model
    \item the comparison of the $\Lambda$CDM model with the model with dark energy parameterised as 2b is inconclusive, a more accurate data set is required to distinguish those models;
    \item damping term, i.e. $(1+z)^3$, which appears in parameterisations (3) is
    not supported by the data used in analysis.
\end{itemize}
In study of cosmological constraints on the form of dark energy the most popular
methodology is study of the viability of different parameterisations for the
equation of state parameter. They are postulated rather in the a priori forms
without connection with true model dynamics. Our approach is different because
we claim that if model dynamics is closed then corresponding form of dark energy
parameterisation should be forced. It is because we tested the FRW model with
dark energy rather than the parameterisation $w(z)$ itself.

\begin{appendix}
\section{Linearised formulas for w(z).}
\label{appa}
Here we present linearised formulas for w(z) around the critical point
corresponding to the deSitter state for monotonic and oscillating evolution
toward this point \cite{Hrycyna:2007gd}:
\begin{equation}
  w_{X}^{\mathrm{mon}} = \frac{-(1-3\xi) + f_{1}(\xi,a)a^{-3/2} +
  f_{2}(\xi,a)a^{-3}}
  {(1-3\xi) + 6\xi(1-6\xi)\psi_{0}(A a^{\alpha_{l}}+Ba^{-\alpha_{l}})a^{-3/2} + 
  3\xi(1-6\xi)(A a^{\alpha_{l}}+ B a^{-\alpha_{l}})^{2}a^{-3}},
  \label{lin}
  \end{equation}
where $\psi_{0}^{2}=\frac{1}{6\xi}$, $\alpha_{l}=\frac{\sqrt{3}}{2}\sqrt{\frac{3-25\xi}{1-3\xi}}$, $A =\frac{1}{2}x_{0} +\sqrt{3}\sqrt{\frac{1-3\xi}{3-25\xi}}\Big(\frac{1}{2}x_{0}+\frac{1}{3}y_{0}\Big)$, $B=\frac{1}{2}x_{0}-\sqrt{3}\sqrt{\frac{1-3\xi}{3-25\xi}}\Big(\frac{1}{2}x_{0}+\frac{1}{3}y_{0}\Big)$, $x_{0}$ and $y_{0}$ are the initial conditions for $\psi$ and $\psi'$, respectively, and $f_{1}=2\xi\psi_{0}\Big(\big(3(1-4\xi)-4\alpha_{l}(1-3\xi)\big)Aa^{\alpha_{l}}+\big(3(1-4\xi)+4\alpha_{l}(1-3\xi)\big)Ba^{-\alpha_{l}}\Big)$, $f_{2}=\big(-\frac{3}{4}(3-4\xi)+15\xi(1-2\xi)\big)\Big(A a^{\alpha_{l}}+Ba^{-\alpha_{l}}\Big)^{2} + \alpha_{l}\big(3(1-4\xi)-8\xi(1-3\xi)\big)\Big(A^{2}a^{2\alpha_{l}} -B^{2}a^{-2\alpha_{l}}\Big) -\alpha_{l}^{2}(1-4\xi)\Big(Aa^{\alpha_{l}}-Ba^{-\alpha_{l}}\Big)^{2}$, and
\begin{equation}
      w_{X}^{\mathrm{osc}} =
      \frac{-(1-3\xi)+g_{1}(\xi,a)a^{-3/2}+g_{2}(\xi,a)a^{-3}}
      {(1-3\xi)+6\xi(1-6\xi)\psi_{0}h(\xi,a)a^{-3/2}
      +3\xi(1-6\xi)h^{2}(\xi,a)a^{-3}
      }, 
      \label{osc}
\end{equation}
where $h=x_{0}\cos{(\alpha_{\mathrm{osc}}\ln{a})+\frac{3}{\alpha_{\mathrm{osc}}}
\sin{(\alpha_{\mathrm{osc}}\ln{a})\big(\frac{1}{2}x_{0}+\frac{1}{3}y_{0}\big)}}$, $g_{1}=2\xi\psi_{0}\Big((1-6\xi)h-4(1-3\xi)\big((x_{0}+y_{0})
\cos{(\alpha_{\mathrm{osc}}\ln{a})}-\alpha_{\mathrm{osc}}x_{0}\sin{(\alpha_
{\mathrm{osc}}\ln{a})}-\frac{3}{2\alpha_{\mathrm{osc}}}\sin{(\alpha_{\mathrm{osc}}
\ln{a})(\frac{1}{2}x_{0}+\frac{1}{3}y_{0})}\big)\Big)$, $g_{2}=\xi(1-6\xi)h^{2}-(1-2\xi)(1-6\xi)\big(y_{0}\cos{(\alpha_{\mathrm{osc}}
\ln{a})}-\alpha_{\mathrm{osc}}x_{0}\sin{(\alpha_{\mathrm{osc}}\ln{a})}-\frac{9}
{2\alpha_{\mathrm{osc}}}\sin{(\alpha_{\mathrm{osc}}\ln{a})}(\frac{1}{2}x_{0}
+\frac{1}{3}y_{0})\big)^{2}-4\xi(1-3\xi)\Big((x_{0}+y_{0})\cos{
(\alpha_{\mathrm{osc}}\ln{a})}-\alpha_{\mathrm{osc}}x_{0}\sin{(\alpha_{\mathrm{osc}}
\ln{a})}-\frac{3}{2\alpha_{\mathrm{osc}}}\sin{(\alpha_{\mathrm{osc}}\ln{a})(\frac{1}{2}
x_{0}+\frac{1}{3}y_{0})}\Big)^{2}$, where $\alpha_{\mathrm{osc}}=\frac{\sqrt{3}}{2}\sqrt{\frac{25\xi-3}{1-3\xi}}$ and $x_{0}$, $y_{0}$ and $\psi_{0}$ have their usual meaning.

Note that in all cases purely oscillating scenario does not exist.

\end{appendix}

\begin{acknowledgments}
This work has been supported by the Marie Curie Host Fellowships for the
Transfer of Knowledge project COCOS (Contract No. MTKD-CT-2004-517186). The
authors also acknowledge cooperation in the project
PARTICLE PHYSICS AND COSMOLOGY: THE INTERFACE (Particles-Astrophysics-Cosmology
Agreement for scientific collaboration in theoretical research).
\end{acknowledgments}

\bibliography{parEoS_R1}

\begin{thebibliography}{10}%
\makeatletter
\providecommand \@ifxundefined [1]{%
 \ifx #1\undefined \expandafter \@firstoftwo
 \else \expandafter \@secondoftwo
\fi
}%
\providecommand \@ifnum [1]{%
 \ifnum #1\expandafter \@firstoftwo
 \else \expandafter \@secondoftwo
\fi
}%
\providecommand \enquote [1]{``#1''}%
\providecommand \bibnamefont  [1]{#1}%
\providecommand \bibfnamefont [1]{#1}%
\providecommand \citenamefont [1]{#1}%
\providecommand\href[0]{\@sanitize\@href}%
\providecommand\@href[1]{\endgroup\@@startlink{#1}\endgroup\@@href}%
\providecommand\@@href[1]{#1\@@endlink}%
\providecommand \@sanitize [0]{\begingroup\catcode`\&12\catcode`\#12\relax}%
\@ifxundefined \pdfoutput {\@firstoftwo}{%
 \@ifnum{\z@=\pdfoutput}{\@firstoftwo}{\@secondoftwo}%
}{%
 \providecommand\@@startlink[1]{\leavevmode\special{html:<a href="#1">}}%
 \providecommand\@@endlink[0]{\special{html:</a>}}%
}{%
 \providecommand\@@startlink[1]{%
  \leavevmode
  \pdfstartlink
   attr{/Border[0 0 1 ]/H/I/C[0 1 1]}%
   user{/Subtype/Link/A<</Type/Action/S/URI/URI(#1)>>}%
  \relax
 }%
 \providecommand\@@endlink[0]{\pdfendlink}%
}%
\providecommand \url  [0]{\begingroup\@sanitize \@url }%
\providecommand \@url [1]{\endgroup\@href {#1}{\urlprefix}}%
\providecommand \urlprefix [0]{URL }%
\providecommand \Eprint[0]{\href }%
\@ifxundefined \urlstyle {%
  \providecommand \doi [1]{doi:\discretionary{}{}{}#1}%
}{%
  \providecommand \doi [0]{doi:\discretionary{}{}{}\begingroup
  \urlstyle{rm}\Url }%
}%
\providecommand \doibase [0]{http://dx.doi.org/}%
\providecommand \Doi[1]{\href{\doibase#1}}%
\providecommand \bibAnnote [3]{%
  \BibitemShut{#1}%
  \begin{quotation}\noindent
    \textsc{Key:}\ #2\\\textsc{Annotation:}\ #3%
  \end{quotation}%
}%
\providecommand \bibAnnoteFile [2]{%
  \IfFileExists{#2}{\bibAnnote {#1} {#2} {\input{#2}}}{}%
}%
\providecommand \typeout [0]{\immediate \write \m@ne }%
\providecommand \selectlanguage [0]{\@gobble}%
\providecommand \bibinfo [0]{\@secondoftwo}%
\providecommand \bibfield [0]{\@secondoftwo}%
\providecommand \translation [1]{[#1]}%
\providecommand \BibitemOpen[0]{}%
\providecommand \bibitemStop [0]{}%
\providecommand \bibitemNoStop [0]{.\EOS\space}%
\providecommand \EOS [0]{\spacefactor3000\relax}%
\providecommand \BibitemShut [1]{\csname bibitem#1\endcsname}%
\bibitem{Riess:1998cb}%
  \BibitemOpen
  \bibfield{author}{%
  \bibinfo {author} {\bibfnamefont{A.~G.}\ \bibnamefont{Riess}} \emph{et~al.}
  (\bibinfo {collaboration} {Supernova Search Team}),\ }%
  \bibfield{journal}{%
  \Doi{10.1086/300499}{\bibinfo {journal} {Astron. J.}}\ }%
  \textbf{\bibinfo {volume} {116}},\ \bibinfo {pages} {1009} (\bibinfo {year}
  {1998}),\
  \Eprint{http://arxiv.org/abs/astro-ph/9805201}{arXiv:astro-ph/9805201}%
  \bibAnnoteFile{NoStop}{Riess:1998cb}%
\bibitem{Perlmutter:1998np}%
  \BibitemOpen
  \bibfield{author}{%
  \bibinfo {author} {\bibfnamefont{S.}~\bibnamefont{Perlmutter}} \emph{et~al.}
  (\bibinfo {collaboration} {Supernova Cosmology Project}),\ }%
  \bibfield{journal}{%
  \Doi{10.1086/307221}{\bibinfo {journal} {Astrophys. J.}}\ }%
  \textbf{\bibinfo {volume} {517}},\ \bibinfo {pages} {565} (\bibinfo {year}
  {1999}),\
  \Eprint{http://arxiv.org/abs/astro-ph/9812133}{arXiv:astro-ph/9812133}%
  \bibAnnoteFile{NoStop}{Perlmutter:1998np}%
\bibitem{Bennett:2003bz}%
  \BibitemOpen
  \bibfield{author}{%
  \bibinfo {author} {\bibfnamefont{C.~L.}\ \bibnamefont{Bennett}} \emph{et~al.}
  (\bibinfo {collaboration} {WMAP}),\ }%
  \bibfield{journal}{%
  \Doi{10.1086/377253}{\bibinfo {journal} {Astrophys. J. Suppl.}}\ }%
  \textbf{\bibinfo {volume} {148}},\ \bibinfo {pages} {1} (\bibinfo {year}
  {2003}),\
  \Eprint{http://arxiv.org/abs/astro-ph/0302207}{arXiv:astro-ph/0302207}%
  \bibAnnoteFile{NoStop}{Bennett:2003bz}%
\bibitem{Spergel:2006hy}%
  \BibitemOpen
  \bibfield{author}{%
  \bibinfo {author} {\bibfnamefont{D.~N.}\ \bibnamefont{Spergel}} \emph{et~al.}
  (\bibinfo {collaboration} {WMAP}),\ }%
  \bibfield{journal}{%
  \Doi{10.1086/513700}{\bibinfo {journal} {Astrophys. J. Suppl.}}\ }%
  \textbf{\bibinfo {volume} {170}},\ \bibinfo {pages} {377} (\bibinfo {year}
  {2007}),\
  \Eprint{http://arxiv.org/abs/astro-ph/0603449}{arXiv:astro-ph/0603449}%
  \bibAnnoteFile{NoStop}{Spergel:2006hy}%
\bibitem{Tegmark:2006az}%
  \BibitemOpen
  \bibfield{author}{%
  \bibinfo {author} {\bibfnamefont{M.}~\bibnamefont{Tegmark}} \emph{et~al.}
  (\bibinfo {collaboration} {SDSS}),\ }%
  \bibfield{journal}{%
  \Doi{10.1103/PhysRevD.74.123507}{\bibinfo {journal} {Phys. Rev.}}\ }%
  \textbf{\bibinfo {volume} {D74}},\ \bibinfo {pages} {123507} (\bibinfo {year}
  {2006}),\
  \Eprint{http://arxiv.org/abs/astro-ph/0608632}{arXiv:astro-ph/0608632}%
  \bibAnnoteFile{NoStop}{Tegmark:2006az}%
\bibitem{Szydlowski:2006ay}%
  \BibitemOpen
  \bibfield{author}{%
  \bibinfo {author} {\bibfnamefont{M.}~\bibnamefont{Szydlowski}}, \bibinfo
  {author} {\bibfnamefont{A.}~\bibnamefont{Kurek}},\ and\ \bibinfo {author}
  {\bibfnamefont{A.}~\bibnamefont{Krawiec}},\ }%
  \bibfield{journal}{%
  \Doi{10.1016/j.physletb.2006.09.052}{\bibinfo {journal} {Phys. Lett.}}\ }%
  \textbf{\bibinfo {volume} {B642}},\ \bibinfo {pages} {171} (\bibinfo {year}
  {2006}),\
  \Eprint{http://arxiv.org/abs/astro-ph/0604327}{arXiv:astro-ph/0604327}%
  \bibAnnoteFile{NoStop}{Szydlowski:2006ay}%
\bibitem{Szydlowski:2006pz}%
  \BibitemOpen
  \bibfield{author}{%
  \bibinfo {author} {\bibfnamefont{M.}~\bibnamefont{Szydlowski}}\ and\ \bibinfo
  {author} {\bibfnamefont{A.}~\bibnamefont{Kurek}},\ }%
  \bibfield{journal}{%
  \Doi{10.1063/1.2399695}{\bibinfo {journal} {AIP Conf. Proc.}}\ }%
  \textbf{\bibinfo {volume} {861}},\ \bibinfo {pages} {1031} (\bibinfo {year}
  {2006}),\
  \Eprint{http://arxiv.org/abs/astro-ph/0603538}{arXiv:astro-ph/0603538}%
  \bibAnnoteFile{NoStop}{Szydlowski:2006pz}%
\bibitem{Kurek:2007tb}%
  \BibitemOpen
  \bibfield{author}{%
  \bibinfo {author} {\bibfnamefont{A.}~\bibnamefont{Kurek}}\ and\ \bibinfo
  {author} {\bibfnamefont{M.}~\bibnamefont{Szydlowski}},\ }%
  \bibfield{journal}{%
  \Doi{10.1086/526333}{\bibinfo {journal} {Astrophys. J.}}\ }%
  \textbf{\bibinfo {volume} {675}},\ \bibinfo {pages} {1} (\bibinfo {year}
  {2008}),\
  \Eprint{http://arxiv.org/abs/astro-ph/0702484}{arXiv:astro-ph/0702484}%
  \bibAnnoteFile{NoStop}{Kurek:2007tb}%
\bibitem{Kurek:2007gr}%
  \BibitemOpen
  \bibfield{author}{%
  \bibinfo {author} {\bibfnamefont{A.}~\bibnamefont{Kurek}}\ and\ \bibinfo
  {author} {\bibfnamefont{M.}~\bibnamefont{Szydlowski}},\ }%
  \bibfield{journal}{%
  \Doi{10.1393/ncb/i2008-10480-3}{\bibinfo {journal} {Nuovo Cim.}}\ }%
  \textbf{\bibinfo {volume} {122B}},\ \bibinfo {pages} {1359} (\bibinfo {year}
  {2007}),\ \Eprint{http://arxiv.org/abs/0710.2125}{arXiv:0710.2125
  [astro-ph]}%
  \bibAnnoteFile{NoStop}{Kurek:2007gr}%
\bibitem{Padmanabhan:2002ji}%
  \BibitemOpen
  \bibfield{author}{%
  \bibinfo {author} {\bibfnamefont{T.}~\bibnamefont{Padmanabhan}},\ }%
  \bibfield{journal}{%
  \Doi{10.1016/S0370-1573(03)00120-0}{\bibinfo {journal} {Phys. Rept.}}\ }%
  \textbf{\bibinfo {volume} {380}},\ \bibinfo {pages} {235} (\bibinfo {year}
  {2003}),\ \Eprint{http://arxiv.org/abs/hep-th/0212290}{arXiv:hep-th/0212290}%
  \bibAnnoteFile{NoStop}{Padmanabhan:2002ji}%
\bibitem{Bludman:2006cg}%
  \BibitemOpen
  \bibfield{author}{%
  \bibinfo {author} {\bibfnamefont{S.~A.}\ \bibnamefont{Bludman}},\ }%
  \enquote{\bibinfo {title} {{Cosmological acceleration: Dark energy or
  modified gravity?}}.}\  (\bibinfo {year} {2006}),\
  \Eprint{http://arxiv.org/abs/astro-ph/0605198}{arXiv:astro-ph/0605198}%
  \bibAnnoteFile{NoStop}{Bludman:2006cg}%
\bibitem{Linder:2005dw}%
  \BibitemOpen
  \bibfield{author}{%
  \bibinfo {author} {\bibfnamefont{E.~V.}\ \bibnamefont{Linder}},\ }%
  \bibfield{journal}{%
  \Doi{10.1016/j.astropartphys.2005.12.003}{\bibinfo {journal} {Astropart.
  Phys.}}\ }%
  \textbf{\bibinfo {volume} {25}},\ \bibinfo {pages} {167} (\bibinfo {year}
  {2006}),\
  \Eprint{http://arxiv.org/abs/astro-ph/0511415}{arXiv:astro-ph/0511415}%
  \bibAnnoteFile{NoStop}{Linder:2005dw}%
\bibitem{Xia:2004rw}%
  \BibitemOpen
  \bibfield{author}{%
  \bibinfo {author} {\bibfnamefont{J.-Q.}\ \bibnamefont{Xia}}, \bibinfo
  {author} {\bibfnamefont{B.}~\bibnamefont{Feng}},\ and\ \bibinfo {author}
  {\bibfnamefont{X.-M.}\ \bibnamefont{Zhang}},\ }%
  \bibfield{journal}{%
  \Doi{10.1142/S0217732305017445}{\bibinfo {journal} {Mod. Phys. Lett.}}\ }%
  \textbf{\bibinfo {volume} {A20}},\ \bibinfo {pages} {2409} (\bibinfo {year}
  {2005}),\
  \Eprint{http://arxiv.org/abs/astro-ph/0411501}{arXiv:astro-ph/0411501}%
  \bibAnnoteFile{NoStop}{Xia:2004rw}%
\bibitem{Barenboim:2004kz}%
  \BibitemOpen
  \bibfield{author}{%
  \bibinfo {author} {\bibfnamefont{G.}~\bibnamefont{Barenboim}}, \bibinfo
  {author} {\bibfnamefont{O.}~\bibnamefont{Mena}},\ and\ \bibinfo {author}
  {\bibfnamefont{C.}~\bibnamefont{Quigg}},\ }%
  \bibfield{journal}{%
  \Doi{10.1103/PhysRevD.71.063533}{\bibinfo {journal} {Phys. Rev.}}\ }%
  \textbf{\bibinfo {volume} {D71}},\ \bibinfo {pages} {063533} (\bibinfo {year}
  {2005}),\
  \Eprint{http://arxiv.org/abs/astro-ph/0412010}{arXiv:astro-ph/0412010}%
  \bibAnnoteFile{NoStop}{Barenboim:2004kz}%
\bibitem{Zhao:2006mn}%
  \BibitemOpen
  \bibfield{author}{%
  \bibinfo {author} {\bibfnamefont{W.}~\bibnamefont{Zhao}},\ }%
  \bibfield{journal}{%
  \Doi{10.1088/1009-1963/16/9/056}{\bibinfo {journal} {Chin. Phys.}}\ }%
  \textbf{\bibinfo {volume} {16}},\ \bibinfo {pages} {2830} (\bibinfo {year}
  {2007}),\
  \Eprint{http://arxiv.org/abs/astro-ph/0604459}{arXiv:astro-ph/0604459}%
  \bibAnnoteFile{NoStop}{Zhao:2006mn}%
\bibitem{Feng:2004ff}%
  \BibitemOpen
  \bibfield{author}{%
  \bibinfo {author} {\bibfnamefont{B.}~\bibnamefont{Feng}}, \bibinfo {author}
  {\bibfnamefont{M.}~\bibnamefont{Li}}, \bibinfo {author}
  {\bibfnamefont{Y.-S.}\ \bibnamefont{Piao}},\ and\ \bibinfo {author}
  {\bibfnamefont{X.}~\bibnamefont{Zhang}},\ }%
  \bibfield{journal}{%
  \Doi{10.1016/j.physletb.2006.01.066}{\bibinfo {journal} {Phys. Lett.}}\ }%
  \textbf{\bibinfo {volume} {B634}},\ \bibinfo {pages} {101} (\bibinfo {year}
  {2006}),\
  \Eprint{http://arxiv.org/abs/astro-ph/0407432}{arXiv:astro-ph/0407432}%
  \bibAnnoteFile{NoStop}{Feng:2004ff}%
\bibitem{Jain:2007fa}%
  \BibitemOpen
  \bibfield{author}{%
  \bibinfo {author} {\bibfnamefont{D.}~\bibnamefont{Jain}}, \bibinfo {author}
  {\bibfnamefont{A.}~\bibnamefont{Dev}},\ and\ \bibinfo {author}
  {\bibfnamefont{J.~S.}\ \bibnamefont{Alcaniz}},\ }%
  \bibfield{journal}{%
  \Doi{10.1016/j.physletb.2007.09.023}{\bibinfo {journal} {Phys. Lett.}}\ }%
  \textbf{\bibinfo {volume} {B656}},\ \bibinfo {pages} {15} (\bibinfo {year}
  {2007}),\ \Eprint{http://arxiv.org/abs/0709.4234}{arXiv:0709.4234
  [astro-ph]}%
  \bibAnnoteFile{NoStop}{Jain:2007fa}%
\bibitem{Griest:2002cu}%
  \BibitemOpen
  \bibfield{author}{%
  \bibinfo {author} {\bibfnamefont{K.}~\bibnamefont{Griest}},\ }%
  \bibfield{journal}{%
  \Doi{10.1103/PhysRevD.66.123501}{\bibinfo {journal} {Phys. Rev.}}\ }%
  \textbf{\bibinfo {volume} {D66}},\ \bibinfo {pages} {123501} (\bibinfo {year}
  {2002}),\
  \Eprint{http://arxiv.org/abs/astro-ph/0202052}{arXiv:astro-ph/0202052}%
  \bibAnnoteFile{NoStop}{Griest:2002cu}%
\bibitem{Dodelson:2001fq}%
  \BibitemOpen
  \bibfield{author}{%
  \bibinfo {author} {\bibfnamefont{S.}~\bibnamefont{Dodelson}}, \bibinfo
  {author} {\bibfnamefont{M.}~\bibnamefont{Kaplinghat}},\ and\ \bibinfo
  {author} {\bibfnamefont{E.}~\bibnamefont{Stewart}},\ }%
  \bibfield{journal}{%
  \Doi{10.1103/PhysRevLett.85.5276}{\bibinfo {journal} {Phys. Rev. Lett.}}\ }%
  \textbf{\bibinfo {volume} {85}},\ \bibinfo {pages} {5276} (\bibinfo {year}
  {2000}),\
  \Eprint{http://arxiv.org/abs/astro-ph/0002360}{arXiv:astro-ph/0002360}%
  \bibAnnoteFile{NoStop}{Dodelson:2001fq}%
\bibitem{Crittenden:2007yy}%
  \BibitemOpen
  \bibfield{author}{%
  \bibinfo {author} {\bibfnamefont{R.}~\bibnamefont{Crittenden}}, \bibinfo
  {author} {\bibfnamefont{E.}~\bibnamefont{Majerotto}},\ and\ \bibinfo {author}
  {\bibfnamefont{F.}~\bibnamefont{Piazza}},\ }%
  \bibfield{journal}{%
  \Doi{10.1103/PhysRevLett.98.251301}{\bibinfo {journal} {Phys. Rev. Lett.}}\
  }%
  \textbf{\bibinfo {volume} {98}},\ \bibinfo {pages} {251301} (\bibinfo {year}
  {2007}),\
  \Eprint{http://arxiv.org/abs/astro-ph/0702003}{arXiv:astro-ph/0702003}%
  \bibAnnoteFile{NoStop}{Crittenden:2007yy}%
\bibitem{Hrycyna:2007mq}%
  \BibitemOpen
  \bibfield{author}{%
  \bibinfo {author} {\bibfnamefont{O.}~\bibnamefont{Hrycyna}}\ and\ \bibinfo
  {author} {\bibfnamefont{M.}~\bibnamefont{Szydlowski}},\ }%
  \bibfield{journal}{%
  \Doi{10.1016/j.physletb.2007.05.057}{\bibinfo {journal} {Phys. Lett.}}\ }%
  \textbf{\bibinfo {volume} {B651}},\ \bibinfo {pages} {8} (\bibinfo {year}
  {2007}),\ \Eprint{http://arxiv.org/abs/0704.1651}{arXiv:0704.1651 [hep-th]}%
  \bibAnnoteFile{NoStop}{Hrycyna:2007mq}%
\bibitem{Hrycyna:2007gd}%
  \BibitemOpen
  \bibfield{author}{%
  \bibinfo {author} {\bibfnamefont{O.}~\bibnamefont{Hrycyna}}\ and\ \bibinfo
  {author} {\bibfnamefont{M.}~\bibnamefont{Szydlowski}},\ }%
  \bibfield{journal}{%
  \Doi{10.1103/PhysRevD.76.123510}{\bibinfo {journal} {Phys. Rev.}}\ }%
  \textbf{\bibinfo {volume} {D76}},\ \bibinfo {pages} {123510} (\bibinfo {year}
  {2007}),\ \Eprint{http://arxiv.org/abs/0707.4471}{arXiv:0707.4471 [hep-th]}%
  \bibAnnoteFile{NoStop}{Hrycyna:2007gd}%
\bibitem{Kurek:2007bu}%
  \BibitemOpen
  \bibfield{author}{%
  \bibinfo {author} {\bibfnamefont{A.}~\bibnamefont{Kurek}}, \bibinfo {author}
  {\bibfnamefont{O.}~\bibnamefont{Hrycyna}},\ and\ \bibinfo {author}
  {\bibfnamefont{M.}~\bibnamefont{Szydlowski}},\ }%
  \bibfield{journal}{%
  \Doi{10.1016/j.physletb.2007.10.074}{\bibinfo {journal} {Phys. Lett.}}\ }%
  \textbf{\bibinfo {volume} {B659}},\ \bibinfo {pages} {14} (\bibinfo {year}
  {2008}),\ \Eprint{http://arxiv.org/abs/0707.0292}{arXiv:0707.0292
  [astro-ph]}%
  \bibAnnoteFile{NoStop}{Kurek:2007bu}%
\bibitem{Faraoni:2000vg}%
  \BibitemOpen
  \bibfield{author}{%
  \bibinfo {author} {\bibfnamefont{V.}~\bibnamefont{Faraoni}},\ }%
  \bibfield{journal}{%
  \Doi{10.1119/1.1290250}{\bibinfo {journal} {Am. J. Phys.}}\ }%
  \textbf{\bibinfo {volume} {69}},\ \bibinfo {pages} {372} (\bibinfo {year}
  {2001}),\
  \Eprint{http://arxiv.org/abs/physics/0006030}{arXiv:physics/0006030}%
  \bibAnnoteFile{NoStop}{Faraoni:2000vg}%
\bibitem{Ratra:1987rm}%
  \BibitemOpen
  \bibfield{author}{%
  \bibinfo {author} {\bibfnamefont{B.}~\bibnamefont{Ratra}}\ and\ \bibinfo
  {author} {\bibfnamefont{P.~J.~E.}\ \bibnamefont{Peebles}},\ }%
  \bibfield{journal}{%
  \Doi{10.1103/PhysRevD.37.3406}{\bibinfo {journal} {Phys. Rev.}}\ }%
  \textbf{\bibinfo {volume} {D37}},\ \bibinfo {pages} {3406} (\bibinfo {year}
  {1988})%
  \bibAnnoteFile{NoStop}{Ratra:1987rm}%
\bibitem{Wetterich:1987fm}%
  \BibitemOpen
  \bibfield{author}{%
  \bibinfo {author} {\bibfnamefont{C.}~\bibnamefont{Wetterich}},\ }%
  \bibfield{journal}{%
  \Doi{10.1016/0550-3213(88)90193-9}{\bibinfo {journal} {Nucl. Phys.}}\ }%
  \textbf{\bibinfo {volume} {B302}},\ \bibinfo {pages} {668} (\bibinfo {year}
  {1988})%
  \bibAnnoteFile{NoStop}{Wetterich:1987fm}%
\bibitem{Dutta:2008px}%
  \BibitemOpen
  \bibfield{author}{%
  \bibinfo {author} {\bibfnamefont{S.}~\bibnamefont{Dutta}}\ and\ \bibinfo
  {author} {\bibfnamefont{R.~J.}\ \bibnamefont{Scherrer}},\ }%
  \bibfield{journal}{%
  \Doi{10.1103/PhysRevD.78.083512}{\bibinfo {journal} {Phys. Rev.}}\ }%
  \textbf{\bibinfo {volume} {D78}},\ \bibinfo {pages} {083512} (\bibinfo {year}
  {2008}),\ \Eprint{http://arxiv.org/abs/0805.0763}{arXiv:0805.0763
  [astro-ph]}%
  \bibAnnoteFile{NoStop}{Dutta:2008px}%
\bibitem{Johnson:2008se}%
  \BibitemOpen
  \bibfield{author}{%
  \bibinfo {author} {\bibfnamefont{M.~C.}\ \bibnamefont{Johnson}}\ and\
  \bibinfo {author} {\bibfnamefont{M.}~\bibnamefont{Kamionkowski}},\ }%
  \bibfield{journal}{%
  \Doi{10.1103/PhysRevD.78.063010}{\bibinfo {journal} {Phys. Rev.}}\ }%
  \textbf{\bibinfo {volume} {D78}},\ \bibinfo {pages} {063010} (\bibinfo {year}
  {2008}),\ \Eprint{http://arxiv.org/abs/0805.1748}{arXiv:0805.1748
  [astro-ph]}%
  \bibAnnoteFile{NoStop}{Johnson:2008se}%
\bibitem{Gu:2007be}%
  \BibitemOpen
  \bibfield{author}{%
  \bibinfo {author} {\bibfnamefont{J.-a.}\ \bibnamefont{Gu}},\ }%
  \enquote{\bibinfo {title} {{Oscillating Quintessence}},}\  (\bibinfo {year}
  {2007}),\ \Eprint{http://arxiv.org/abs/0711.3606}{arXiv:0711.3606
  [astro-ph]}%
  \bibAnnoteFile{NoStop}{Gu:2007be}%
\bibitem{Amendola:1999dr}%
  \BibitemOpen
  \bibfield{author}{%
  \bibinfo {author} {\bibfnamefont{L.}~\bibnamefont{Amendola}},\ }%
  \bibfield{journal}{%
  \Doi{10.1046/j.1365-8711.2000.03165.x}{\bibinfo {journal} {Mon. Not. Roy.
  Astron. Soc.}}\ }%
  \textbf{\bibinfo {volume} {312}},\ \bibinfo {pages} {521} (\bibinfo {year}
  {2000}),\
  \Eprint{http://arxiv.org/abs/astro-ph/9906073}{arXiv:astro-ph/9906073}%
  \bibAnnoteFile{NoStop}{Amendola:1999dr}%
\bibitem{Abbott:1981rg}%
  \BibitemOpen
  \bibfield{author}{%
  \bibinfo {author} {\bibfnamefont{L.~F.}\ \bibnamefont{Abbott}},\ }%
  \bibfield{journal}{%
  \Doi{10.1016/0550-3213(81)90374-6}{\bibinfo {journal} {Nucl. Phys.}}\ }%
  \textbf{\bibinfo {volume} {B185}},\ \bibinfo {pages} {233} (\bibinfo {year}
  {1981})%
  \bibAnnoteFile{NoStop}{Abbott:1981rg}%
\bibitem{MersiniHoughton:2001su}%
  \BibitemOpen
  \bibfield{author}{%
  \bibinfo {author} {\bibfnamefont{L.}~\bibnamefont{Mersini-Houghton}},
  \bibinfo {author} {\bibfnamefont{M.}~\bibnamefont{Bastero-Gil}},\ and\
  \bibinfo {author} {\bibfnamefont{P.}~\bibnamefont{Kanti}},\ }%
  \bibfield{journal}{%
  \Doi{10.1103/PhysRevD.64.043508}{\bibinfo {journal} {Phys. Rev.}}\ }%
  \textbf{\bibinfo {volume} {D64}},\ \bibinfo {pages} {043508} (\bibinfo {year}
  {2001}),\ \Eprint{http://arxiv.org/abs/hep-ph/0101210}{arXiv:hep-ph/0101210}%
  \bibAnnoteFile{NoStop}{MersiniHoughton:2001su}%
\bibitem{Faraoni:2006ik}%
  \BibitemOpen
  \bibfield{author}{%
  \bibinfo {author} {\bibfnamefont{V.}~\bibnamefont{Faraoni}}\ and\ \bibinfo
  {author} {\bibfnamefont{M.~N.}\ \bibnamefont{Jensen}},\ }%
  \bibfield{journal}{%
  \Doi{10.1088/0264-9381/23/9/014}{\bibinfo {journal} {Class. Quant. Grav.}}\
  }%
  \textbf{\bibinfo {volume} {23}},\ \bibinfo {pages} {3005} (\bibinfo {year}
  {2006}),\ \Eprint{http://arxiv.org/abs/gr-qc/0602097}{arXiv:gr-qc/0602097}%
  \bibAnnoteFile{NoStop}{Faraoni:2006ik}%
\bibitem{Faraoni:2005gg}%
  \BibitemOpen
  \bibfield{author}{%
  \bibinfo {author} {\bibfnamefont{V.}~\bibnamefont{Faraoni}},\ }%
  \bibfield{journal}{%
  \Doi{10.1088/0264-9381/22/16/008}{\bibinfo {journal} {Class. Quant. Grav.}}\
  }%
  \textbf{\bibinfo {volume} {22}},\ \bibinfo {pages} {3235} (\bibinfo {year}
  {2005}),\ \Eprint{http://arxiv.org/abs/gr-qc/0506095}{arXiv:gr-qc/0506095}%
  \bibAnnoteFile{NoStop}{Faraoni:2005gg}%
\bibitem{Faraoni:2000gx}%
  \BibitemOpen
  \bibfield{author}{%
  \bibinfo {author} {\bibfnamefont{V.}~\bibnamefont{Faraoni}},\ }%
  \bibfield{journal}{%
  \Doi{10.1023/A:1012990305341}{\bibinfo {journal} {Int. J. Theor. Phys.}}\ }%
  \textbf{\bibinfo {volume} {40}},\ \bibinfo {pages} {2259} (\bibinfo {year}
  {2001}),\ \Eprint{http://arxiv.org/abs/hep-th/0009053}{arXiv:hep-th/0009053}%
  \bibAnnoteFile{NoStop}{Faraoni:2000gx}%
\bibitem{Szydlowski:2008zza}%
  \BibitemOpen
  \bibfield{author}{%
  \bibinfo {author} {\bibfnamefont{M.}~\bibnamefont{Szydlowski}}, \bibinfo
  {author} {\bibfnamefont{O.}~\bibnamefont{Hrycyna}},\ and\ \bibinfo {author}
  {\bibfnamefont{A.}~\bibnamefont{Kurek}},\ }%
  \bibfield{journal}{%
  \Doi{10.1103/PhysRevD.77.027302}{\bibinfo {journal} {Phys. Rev.}}\ }%
  \textbf{\bibinfo {volume} {D77}},\ \bibinfo {pages} {027302} (\bibinfo {year}
  {2008}),\ \Eprint{http://arxiv.org/abs/0710.0366}{arXiv:0710.0366
  [astro-ph]}%
  \bibAnnoteFile{NoStop}{Szydlowski:2008zza}%
\bibitem{Chevallier:2000qy}%
  \BibitemOpen
  \bibfield{author}{%
  \bibinfo {author} {\bibfnamefont{M.}~\bibnamefont{Chevallier}}\ and\ \bibinfo
  {author} {\bibfnamefont{D.}~\bibnamefont{Polarski}},\ }%
  \bibfield{journal}{%
  \Doi{10.1142/S0218271801000822}{\bibinfo {journal} {Int. J. Mod. Phys.}}\ }%
  \textbf{\bibinfo {volume} {D10}},\ \bibinfo {pages} {213} (\bibinfo {year}
  {2001}),\ \Eprint{http://arxiv.org/abs/gr-qc/0009008}{arXiv:gr-qc/0009008}%
  \bibAnnoteFile{NoStop}{Chevallier:2000qy}%
\bibitem{Linder:2004ng}%
  \BibitemOpen
  \bibfield{author}{%
  \bibinfo {author} {\bibfnamefont{E.~V.}\ \bibnamefont{Linder}},\ }%
  \bibfield{journal}{%
  \Doi{10.1103/PhysRevD.70.023511}{\bibinfo {journal} {Phys. Rev.}}\ }%
  \textbf{\bibinfo {volume} {D70}},\ \bibinfo {pages} {023511} (\bibinfo {year}
  {2004}),\
  \Eprint{http://arxiv.org/abs/astro-ph/0402503}{arXiv:astro-ph/0402503}%
  \bibAnnoteFile{NoStop}{Linder:2004ng}%
\bibitem{Jeffreys:1961}%
  \BibitemOpen
  \bibfield{author}{%
  \bibinfo {author} {\bibfnamefont{H.}~\bibnamefont{Jeffreys}},\ }%
  \emph{\bibinfo {title} {Theory of Probability}},\ \bibinfo {edition} {3rd}\
  ed.\ (\bibinfo {publisher} {Oxford University Press},\ \bibinfo {address}
  {Oxford},\ \bibinfo {year} {1961})%
  \bibAnnoteFile{NoStop}{Jeffreys:1961}%
\bibitem{Trotta:2008qt}%
  \BibitemOpen
  \bibfield{author}{%
  \bibinfo {author} {\bibfnamefont{R.}~\bibnamefont{Trotta}},\ }%
  \bibfield{journal}{%
  \bibinfo {journal} {Contemp. Phys.}\ }%
  \textbf{\bibinfo {volume} {49}},\ \bibinfo {pages} {71} (\bibinfo {year}
  {2008}),\ \Eprint{http://arxiv.org/abs/0803.4089}{arXiv:0803.4089
  [astro-ph]}%
  \bibAnnoteFile{NoStop}{Trotta:2008qt}%
\bibitem{Skilling}%
  \BibitemOpen
  \bibfield{author}{%
  \bibinfo {author} {\bibfnamefont{J.}~\bibnamefont{Skilling}},\ }%
  \bibinfo {note} {http://www.inference.phy.cam.ac.uk/bayesys/}%
  \bibAnnoteFile{NoStop}{Skilling}%
\bibitem{cosmo:1}%
  \BibitemOpen
  \bibfield{author}{%
  \bibinfo {author} {\bibfnamefont{A.}~\bibnamefont{Lewis}},\ }%
  \enquote{\bibinfo {title} {{CosmoMC}},}\ \bibinfo {note}
  {\url{http://cosmologist.info/}}%
  \bibAnnoteFile{NoStop}{cosmo:1}%
\bibitem{Lewis:2002ah}%
  \BibitemOpen
  \bibfield{author}{%
  \bibinfo {author} {\bibfnamefont{A.}~\bibnamefont{Lewis}}\ and\ \bibinfo
  {author} {\bibfnamefont{S.}~\bibnamefont{Bridle}},\ }%
  \bibfield{journal}{%
  \bibinfo {journal} {Phys. Rev.}\ }%
  \textbf{\bibinfo {volume} {D66}},\ \bibinfo {pages} {103511} (\bibinfo {year}
  {2002}),\ \Eprint{http://arxiv.org/abs/astro-ph/0205436}{astro-ph/0205436}%
  \bibAnnoteFile{NoStop}{Lewis:2002ah}%
\bibitem{cosmo:2}%
  \BibitemOpen
  \bibfield{author}{%
  \bibinfo {author} {\bibfnamefont{P.~M.}\ \bibnamefont{D.~Parkinson}}\ and\
  \bibinfo {author} {\bibfnamefont{A.}~\bibnamefont{Liddle}},\ }%
  \enquote{\bibinfo {title} {Cosmonest},}\ \bibinfo {note}
  {Http://astronomy.sussex.ac.uk/~pm52/cosmonest/}%
  \bibAnnoteFile{NoStop}{cosmo:2}%
\bibitem{Mukherjee:2005wg}%
  \BibitemOpen
  \bibfield{author}{%
  \bibinfo {author} {\bibfnamefont{P.}~\bibnamefont{Mukherjee}}, \bibinfo
  {author} {\bibfnamefont{D.}~\bibnamefont{Parkinson}},\ and\ \bibinfo {author}
  {\bibfnamefont{A.~R.}\ \bibnamefont{Liddle}},\ }%
  \bibfield{journal}{%
  \bibinfo {journal} {Astrophys. J.}\ }%
  \textbf{\bibinfo {volume} {638}},\ \bibinfo {pages} {L51} (\bibinfo {year}
  {2006}),\
  \Eprint{http://arxiv.org/abs/astro-ph/0508461}{arXiv:astro-ph/0508461}%
  \bibAnnoteFile{NoStop}{Mukherjee:2005wg}%
\bibitem{Mukherjee:2005tr}%
  \BibitemOpen
  \bibfield{author}{%
  \bibinfo {author} {\bibfnamefont{P.}~\bibnamefont{Mukherjee}}, \bibinfo
  {author} {\bibfnamefont{D.}~\bibnamefont{Parkinson}}, \bibinfo {author}
  {\bibfnamefont{P.~S.}\ \bibnamefont{Corasaniti}}, \bibinfo {author}
  {\bibfnamefont{A.~R.}\ \bibnamefont{Liddle}},\ and\ \bibinfo {author}
  {\bibfnamefont{M.}~\bibnamefont{Kunz}},\ }%
  \bibfield{journal}{%
  \Doi{10.1111/j.1365-2966.2006.10427.x}{\bibinfo {journal} {Mon. Not. Roy.
  Astron. Soc.}}\ }%
  \textbf{\bibinfo {volume} {369}},\ \bibinfo {pages} {1725} (\bibinfo {year}
  {2006}),\
  \Eprint{http://arxiv.org/abs/astro-ph/0512484}{arXiv:astro-ph/0512484}%
  \bibAnnoteFile{NoStop}{Mukherjee:2005tr}%
\bibitem{Parkinson:2006ku}%
  \BibitemOpen
  \bibfield{author}{%
  \bibinfo {author} {\bibfnamefont{D.}~\bibnamefont{Parkinson}}, \bibinfo
  {author} {\bibfnamefont{P.}~\bibnamefont{Mukherjee}},\ and\ \bibinfo {author}
  {\bibfnamefont{A.~R.}\ \bibnamefont{Liddle}},\ }%
  \bibfield{journal}{%
  \Doi{10.1103/PhysRevD.73.123523}{\bibinfo {journal} {Phys. Rev.}}\ }%
  \textbf{\bibinfo {volume} {D73}},\ \bibinfo {pages} {123523} (\bibinfo {year}
  {2006}),\
  \Eprint{http://arxiv.org/abs/astro-ph/0605003}{arXiv:astro-ph/0605003}%
  \bibAnnoteFile{NoStop}{Parkinson:2006ku}%
\bibitem{Davis:2007na}%
  \BibitemOpen
  \bibfield{author}{%
  \bibinfo {author} {\bibfnamefont{T.~M.}\ \bibnamefont{Davis}} \emph{et~al.},\
  }%
  \bibfield{journal}{%
  \Doi{10.1086/519988}{\bibinfo {journal} {Astrophys. J.}}\ }%
  \textbf{\bibinfo {volume} {666}},\ \bibinfo {pages} {716} (\bibinfo {year}
  {2007}),\
  \Eprint{http://arxiv.org/abs/astro-ph/0701510}{arXiv:astro-ph/0701510}%
  \bibAnnoteFile{NoStop}{Davis:2007na}%
\bibitem{WoodVasey:2007jb}%
  \BibitemOpen
  \bibfield{author}{%
  \bibinfo {author} {\bibfnamefont{W.~M.}\ \bibnamefont{Wood-Vasey}}
  \emph{et~al.} (\bibinfo {collaboration} {ESSENCE}),\ }%
  \bibfield{journal}{%
  \Doi{10.1086/518642}{\bibinfo {journal} {Astrophys. J.}}\ }%
  \textbf{\bibinfo {volume} {666}},\ \bibinfo {pages} {694} (\bibinfo {year}
  {2007}),\
  \Eprint{http://arxiv.org/abs/astro-ph/0701041}{arXiv:astro-ph/0701041}%
  \bibAnnoteFile{NoStop}{WoodVasey:2007jb}%
\bibitem{Riess:2006fw}%
  \BibitemOpen
  \bibfield{author}{%
  \bibinfo {author} {\bibfnamefont{A.~G.}\ \bibnamefont{Riess}} \emph{et~al.},\
  }%
  \bibfield{journal}{%
  \Doi{10.1086/510378}{\bibinfo {journal} {Astrophys. J.}}\ }%
  \textbf{\bibinfo {volume} {659}},\ \bibinfo {pages} {98} (\bibinfo {year}
  {2007}),\
  \Eprint{http://arxiv.org/abs/astro-ph/0611572}{arXiv:astro-ph/0611572}%
  \bibAnnoteFile{NoStop}{Riess:2006fw}%
\bibitem{Wang:2006ts}%
  \BibitemOpen
  \bibfield{author}{%
  \bibinfo {author} {\bibfnamefont{Y.}~\bibnamefont{Wang}}\ and\ \bibinfo
  {author} {\bibfnamefont{P.}~\bibnamefont{Mukherjee}},\ }%
  \bibfield{journal}{%
  \Doi{10.1086/507091}{\bibinfo {journal} {Astrophys. J.}}\ }%
  \textbf{\bibinfo {volume} {650}},\ \bibinfo {pages} {1} (\bibinfo {year}
  {2006}),\
  \Eprint{http://arxiv.org/abs/astro-ph/0604051}{arXiv:astro-ph/0604051}%
  \bibAnnoteFile{NoStop}{Wang:2006ts}%
\bibitem{Eisenstein:2005su}%
  \BibitemOpen
  \bibfield{author}{%
  \bibinfo {author} {\bibfnamefont{D.~J.}\ \bibnamefont{Eisenstein}}
  \emph{et~al.} (\bibinfo {collaboration} {SDSS}),\ }%
  \bibfield{journal}{%
  \Doi{10.1086/466512}{\bibinfo {journal} {Astrophys. J.}}\ }%
  \textbf{\bibinfo {volume} {633}},\ \bibinfo {pages} {560} (\bibinfo {year}
  {2005}),\
  \Eprint{http://arxiv.org/abs/astro-ph/0501171}{arXiv:astro-ph/0501171}%
  \bibAnnoteFile{NoStop}{Eisenstein:2005su}%
\bibitem{Simon:2004tf}%
  \BibitemOpen
  \bibfield{author}{%
  \bibinfo {author} {\bibfnamefont{J.}~\bibnamefont{Simon}}, \bibinfo {author}
  {\bibfnamefont{L.}~\bibnamefont{Verde}},\ and\ \bibinfo {author}
  {\bibfnamefont{R.}~\bibnamefont{Jimenez}},\ }%
  \bibfield{journal}{%
  \Doi{10.1103/PhysRevD.71.123001}{\bibinfo {journal} {Phys. Rev.}}\ }%
  \textbf{\bibinfo {volume} {D71}},\ \bibinfo {pages} {123001} (\bibinfo {year}
  {2005}),\
  \Eprint{http://arxiv.org/abs/astro-ph/0412269}{arXiv:astro-ph/0412269}%
  \bibAnnoteFile{NoStop}{Simon:2004tf}%
\bibitem{Gaztanaga:2008xz}%
  \BibitemOpen
  \bibfield{author}{%
  \bibinfo {author} {\bibfnamefont{E.}~\bibnamefont{Gaztanaga}}, \bibinfo
  {author} {\bibfnamefont{A.}~\bibnamefont{Cabre}},\ and\ \bibinfo {author}
  {\bibfnamefont{L.}~\bibnamefont{Hui}},\ }%
  \bibfield{journal}{%
  \Doi{10.1111/j.1365-2966.2009.15405.x}{\bibinfo {journal} {Mon. Not. Roy.
  Astron. Soc.}}\ }%
  \textbf{\bibinfo {volume} {399}},\ \bibinfo {pages} {1663} (\bibinfo {year}
  {2009}),\ \Eprint{http://arxiv.org/abs/0807.3551}{arXiv:0807.3551
  [astro-ph]}%
  \bibAnnoteFile{NoStop}{Gaztanaga:2008xz}%
\bibitem{Freedman:2000cf}%
  \BibitemOpen
  \bibfield{author}{%
  \bibinfo {author} {\bibfnamefont{W.~L.}\ \bibnamefont{Freedman}}
  \emph{et~al.} (\bibinfo {collaboration} {HST}),\ }%
  \bibfield{journal}{%
  \Doi{10.1086/320638}{\bibinfo {journal} {Astrophys. J.}}\ }%
  \textbf{\bibinfo {volume} {553}},\ \bibinfo {pages} {47} (\bibinfo {year}
  {2001}),\
  \Eprint{http://arxiv.org/abs/astro-ph/0012376}{arXiv:astro-ph/0012376}%
  \bibAnnoteFile{NoStop}{Freedman:2000cf}%
\bibitem{Wei:2008rv}%
  \BibitemOpen
  \bibfield{author}{%
  \bibinfo {author} {\bibfnamefont{H.}~\bibnamefont{Wei}},\ }%
  \bibfield{journal}{%
  \Doi{10.1140/epjc/s10052-009-0891-8}{\bibinfo {journal} {Eur. Phys. J.}}\ }%
  \textbf{\bibinfo {volume} {C60}},\ \bibinfo {pages} {449} (\bibinfo {year}
  {2009}),\ \Eprint{http://arxiv.org/abs/0809.0057}{arXiv:0809.0057
  [astro-ph]}%
  \bibAnnoteFile{NoStop}{Wei:2008rv}%
\bibitem{Hawkins:2002sg}%
  \BibitemOpen
  \bibfield{author}{%
  \bibinfo {author} {\bibfnamefont{E.}~\bibnamefont{Hawkins}} \emph{et~al.},\
  }%
  \bibfield{journal}{%
  \Doi{10.1046/j.1365-2966.2003.07063.x}{\bibinfo {journal} {Mon. Not. Roy.
  Astron. Soc.}}\ }%
  \textbf{\bibinfo {volume} {346}},\ \bibinfo {pages} {78} (\bibinfo {year}
  {2003}),\
  \Eprint{http://arxiv.org/abs/astro-ph/0212375}{arXiv:astro-ph/0212375}%
  \bibAnnoteFile{NoStop}{Hawkins:2002sg}%
\bibitem{Verde:2001sf}%
  \BibitemOpen
  \bibfield{author}{%
  \bibinfo {author} {\bibfnamefont{L.}~\bibnamefont{Verde}} \emph{et~al.},\ }%
  \bibfield{journal}{%
  \Doi{10.1046/j.1365-8711.2002.05620.x}{\bibinfo {journal} {Mon. Not. Roy.
  Astron. Soc.}}\ }%
  \textbf{\bibinfo {volume} {335}},\ \bibinfo {pages} {432} (\bibinfo {year}
  {2002}),\
  \Eprint{http://arxiv.org/abs/astro-ph/0112161}{arXiv:astro-ph/0112161}%
  \bibAnnoteFile{NoStop}{Verde:2001sf}%
\bibitem{Guzzo:2008ac}%
  \BibitemOpen
  \bibfield{author}{%
  \bibinfo {author} {\bibfnamefont{L.}~\bibnamefont{Guzzo}} \emph{et~al.},\ }%
  \bibfield{journal}{%
  \Doi{10.1038/nature06555}{\bibinfo {journal} {Nature}}\ }%
  \textbf{\bibinfo {volume} {451}},\ \bibinfo {pages} {541} (\bibinfo {year}
  {2008}),\ \Eprint{http://arxiv.org/abs/0802.1944}{arXiv:0802.1944
  [astro-ph]}%
  \bibAnnoteFile{NoStop}{Guzzo:2008ac}%
\bibitem{McDonald:2004xn}%
  \BibitemOpen
  \bibfield{author}{%
  \bibinfo {author} {\bibfnamefont{P.}~\bibnamefont{McDonald}} \emph{et~al.}
  (\bibinfo {collaboration} {SDSS}),\ }%
  \bibfield{journal}{%
  \Doi{10.1086/497563}{\bibinfo {journal} {Astrophys. J.}}\ }%
  \textbf{\bibinfo {volume} {635}},\ \bibinfo {pages} {761} (\bibinfo {year}
  {2005}),\
  \Eprint{http://arxiv.org/abs/astro-ph/0407377}{arXiv:astro-ph/0407377}%
  \bibAnnoteFile{NoStop}{McDonald:2004xn}%
\end{thebibliography}%
\bibliographystyle{apsrev4-1}

\end{document}